\documentstyle[preprint,prl,aps]{revtex}
\begin{document}

\draft

\title{THE CONTRIBUTION OF 
THE SMECTIC-NEMATIC INTERFACE TO THE SURFACE ENERGY}
\author{L. R. Evangelista, S. Fontanini, L. C. Malacarne and R. S. Mendes}
\address{Departamento de F\'\i sica - Universidade Estadual de Maring\'a \\
Av. Colombo 5790, 87020-900 Maring\'a, Paran\'a, Brazil}
\date{\today}
\maketitle

\begin{abstract}
The contribution of the smectic-nematic interface to the surface energy  of
a nematic liquid crystal sample is analyzed. By means of a simple  model it
is shown that the surface energy depends on the thickness of  the region
over which the transition smectic-nematic takes place. For  perfectly flat
substrates this thickness is of the order of the  correlation length
entering in the transition. An estimate of this  contribution shows that it
is greater than the one arising from  the nematic-substrate interaction.
Moreover, it is also shown that the surface  energy determined in this way
presents a non-monotonic behavior with the  temperature.
\end{abstract}

\date{\today}

The nematic liquid crystalline phase is constituted of long molecules which
present a shape anisotropy. In the nematic mesophase, these molecules tend
to align parallel one to the other thus presenting an orientational order.
The average molecular orientation is described by the unit vector $\vec{n}$,
called the director\cite{de Gennes}. When this system is limited by a flat
surface, the translational invariance of the phase is broken. The presence
of a surface is responsible for an excess of free energy, usually called
surface energy, which has an anisotropic part resulting from the
orientational order characteristic of the nematic phase\cite{Cognard}. This
quantity has been measured by different techniques\cite
{Cognard,Jerome,Blinov,Faetti} and its origin is usually connected to the
nematic-substrate interaction and to the incomplete intermolecular
interaction.

However, the presence of a substrate can also induce a positional order in
the molecules in the vicinity of the surface. In fact, near to the surface
the center of mass of the molecules tend to form layers parallel to the
surface as indicated by several experiments\cite{Jerome}. Since these
molecules are oriented perpendicular or tilted with respect to the layers,
they form smectic layers near the boundaries. As pointed out by Cognard, the
energy confined in these layers is higher than can be added to the LC film
by other actions, like mechanical, thermal or electrical ones\cite{Cognard}.
Therefore, it seems very important to analyze the contribution to the
surface energy coming from the smectic-nematic interface.

The aim of this paper is then to show that the main contribution to the
surface energy of the system can be connected to the interaction between the
smectic layer and the nematic. The  nematic-smectic interface is supposed to
have a thickness $\varepsilon $ along which the system passes from one phase
to the other. On the other hand,  the system can be treated as a simple
junction\cite{Anca}, 
then we explicitly evaluate the surface energy in
the transition zone by introducing the smectic coherence length in the
nematic phase. In this manner also the temperature dependence of
the interfacial energy can be analyzed as is done for the nematic liquid
crystal wall-interface\cite{Rosenblatt}.

Let us consider a nematic slab of thickness $d$. The Cartesian reference
frame has the ($x,y$)-plane coinciding with the surfaces of the slab. The
problem will be supposed one-dimensional with all the quantities depending
only on the $z$-coordinate. The director is everywhere 
parallel to the ($x,z$)-plane and $\phi(z)= \arccos (\vec{n}\cdot \vec{z})$ is the tilt angle
formed by $\vec{n}$ and the $z$-axis. The tilt angle assumes the value 
$\phi_1(z)$ in the smectic layer, $\phi_2(z)$ in the nematic phase and 
$\phi_3(z)$ in the transition region smectic-nematic. By considering that the
transition occurs in a layer of thickness $\varepsilon$ the total elastic
energy per unit surface can be written as

\begin{eqnarray}  \label{1}
F = \int_0^b \frac{1}{2} K_1 {\phi^{\prime}_1}^2 {\rm {d}} z +
\int_b^{b+\varepsilon} \frac{1}{2} K(z) {\phi^{\prime}_3}^2 {\rm {d}} z +
\int_{b+\varepsilon}^{d} \frac{1}{2} K_2 {\phi^{\prime}_2}^2 {\rm {d}} z,
\end{eqnarray}

\noindent where $b$ is the thickness of the smectic region, $K_1$ and $K_2$
are the elastic constants of the smectic and nematic phases respectively,
and $\phi'= d\phi/dz$.
Eq. (\ref{1}) holds in the hypothesis 
that the smectic layer is present only at the
interface close to the surface at $z=0$. A smectic layer can also be 
formed at
$z=d$ interface. However, in order to estimate the contribution of a
smectic-nematic interface to the surface energy, it is sufficient to
consider only the $z=0$ interface.
The second addendum in 
(\ref{1}) represents the contribution to the total elastic energy coming from
the smectic-nematic interface. In this region $K(z)$ can be written, in a
first approximation, as

\begin{eqnarray}  \label{2}
K(z) = K_1 + \frac{K_2-K_1}{\varepsilon} (z-b).
\end{eqnarray}

\noindent  In the strong anchoring hypothesis the boundary conditions at the
surface are $\phi(0)=0$ and $\phi(d)=\Phi$. 
Note that the strong homeotropic anchoring at $z=0$ is equivalent to 
impose the existence of a perfect smectic layer at this border.
To obtain the solutions of the Euler-Lagrange equations resulting from
the minimization of the functional (\ref{1}), we have to consider the continuity
of $\phi(z)$ and of the mechanical torque \cite{Anca} at $z=b$ and $z=b+\epsilon$.
The solutions for each region are

\begin{eqnarray}  \label{5}
\phi_1(z) &=& \frac{C}{K_1} z, \quad 0 \le z \le b,  \nonumber \\
\phi_3(z) &=& \frac{C \varepsilon} {K_2 -K_1} \ln K(z) + C \left[ \frac{b}
{K_1} - \frac{\varepsilon}{K_2-K_1} \ln K_1 \right], \quad b \le z \le
b+\varepsilon,  \nonumber \\
\phi_2(z) &=& \Phi + \frac{C}{K_2} (z-d), \quad b+\varepsilon \le z \le d,
\end{eqnarray}

\noindent where

\begin{eqnarray}  \label{6}
C= \left [ \frac{\varepsilon}{K_2-K_1} \ln\left(\frac{K_2}{K_1}\right) + 
\frac{ d-b-\varepsilon}{K_2} + \frac{b}{K_1} \right ]^{-1} \Phi.
\end{eqnarray}

\noindent From Eqs. (\ref{5}) and (\ref{6}) one obtains for an arbitrary
point $z={\tilde {b}}$

\begin{eqnarray}  \label{7}
\phi_2(\tilde{b}) = \Phi + \frac{C}{K_2} (\tilde{b}-d).
\end{eqnarray}

On the other hand, as already indicated, it is possible to treat the problem
as a smectic-nematic junction where the total elastic energy per unit
surface takes the form

\begin{eqnarray}  \label{8}
{\tilde{F}} = \int_0^{\tilde {b}} \frac{1}{2} K_1 {\tilde {\phi^{\prime}_1}}
^2 {\rm {d}}z + \int_{\tilde{b}}^d \frac{1}{2} K_2 {\tilde {\phi^{\prime}_2}}
^2 {\rm {d}}z + \frac{1}{2} \beta \left[ {\tilde{\phi_1}}({\tilde{b}}) 
- {\tilde{\phi_2}}({\tilde{b}}) \right]^2.
\end{eqnarray}

\noindent In (\ref{8}) the last term represents the contribution connected
to the smectic-nematic junction to the total elastic energy, and $\beta$ is
the surface energy. Again, by minimizing (\ref{8}) subjected to the boundary
conditions ${\tilde{\phi}}(0)=0$ and ${\tilde{\phi}}(d) = \Phi$ one obtains

\begin{eqnarray}  \label{10}
{\tilde{\phi_1}}(z) &=& \frac{{\tilde{C}}}{K_1} z, \quad 0 \le z \le {\tilde{
b}},  \nonumber \\
{\tilde{\phi_2}} (z) &=& \Phi + \frac{{\tilde{C}}}{K_2} (z-d), \quad {\tilde{
b}} \le z \le d,
\end{eqnarray}

\noindent where

\begin{eqnarray}  \label{11}
{\tilde{C}}= \left [ \frac{1}{\beta}+\frac{d-{\tilde{b}}}{K_2} + \frac{{
\tilde{b}}}{K_1} \right ]^{-1} \Phi .
\end{eqnarray}

\noindent From Eq. (\ref{10}) we obtain

\begin{eqnarray}  \label{12}
{\tilde{\phi_2}}({\tilde{b}}) = \Phi + \frac{{\tilde{C}}}{K_2} ({\tilde{b}}
-d).
\end{eqnarray}

The main measurements performed on a real nematic sample concern the bulk
properties, like, for instance, the optical path difference. Since $d \gg b$
and $d \gg \varepsilon$, the physical situations described by $F$ and 
${\tilde{F}}$ must be the same in the bulk. Consequently we will assume that 
$\phi_2(z)={\tilde{\phi_2}}(z)$ and that the border of the nematic phase is
localized in $z=b+\varepsilon$. From Eqs. (\ref{6}), (\ref{7}), (\ref{11})
and (\ref{12}), with ${\tilde{b}}= b+\varepsilon$, we obtain for the surface
energy

\begin{eqnarray}  \label{13}
\beta= \left [ \frac{1}{K_2-K_1} \ln \left( \frac{K_2}{K_1} \right) 
-\frac{1}{K_1} \right ]^{-1} \frac{1}{\varepsilon}.
\end{eqnarray}

\noindent Moreover, if we consider $K_1 = \alpha K_2$ then

\begin{eqnarray}  \label{14}
\beta= \lambda \frac{K_2}{\varepsilon},
\end{eqnarray}

\noindent where

\begin{eqnarray}  \label{15}
\lambda=\frac{\alpha (\alpha -1)} { \alpha \ln \alpha -\alpha +1}.
\end{eqnarray}

\noindent It is important to stress that $\phi_2(z)={\tilde{\phi_2}}(z)$
implies in the equality of the total energies $F$ and ${\tilde{F}}$.

In order to estimate $\beta$ we remember that $\varepsilon$ is essentially
of the order of the correlation length $\xi$ characterizing the region where
the transition smectic-nematic occurs
at a temperature $T^*$. Thus, $\xi$ refers to the coherence length
of the smectic-nematic transition.
Moreover, $\xi$ is expected to be of the
order of several molecular lengths. It is also expected that $\xi$ increases
near the transition temperature. From these considerations a reasonable
estimate is $\varepsilon \approx 1000 \AA$. For a typical nematic like the
PAA, $K_2 \approx 7 \times 10^{-7}\, dyn$ \cite{de Gennes}. Moreover, it is
expected that the elastic constant of the smectic phase, $K_1$ is greater
than the elastic constant of the nematic phase $K_2$. If we assume that 
$\alpha \approx 3$, we obtain for the surface energy, $\beta \approx 0.3 \,
erg/cm^2$. The surface energy measured on real samples are of the order of 
$10^{-1}$ and $10^{-2}\, erg/cm^2$\cite{Blinov}. Therefore our results
indicate that the surface energy is mainly connected to the smectic-nematic
interface, in the hypothesis that $\varepsilon$ is not too large.

If we consider that ${\tilde{b}} < b+\varepsilon$, then, in general, $\beta$
is a negative quantity for $K_1 >K_2$. This situation is not physically
meaningful. It happens only because in this case we are extending the
nematic phase to a region where the phase is not purely nematic.

On the other hand, several recent measurements performed on lyotropic
nematic samples\cite{Sandro} indicates an agreement with the present
estimate. In these experiments with discotic nematic liquid crystals, the
surfaces of the substrates are with and without treatment. The measured
values of the optical path difference 
are the same for both situations. This fact
indicates that the lamellar phase formed between the glass plates and the
nematic liquid crystals is responsible for a strong attenuation of the
effect of the glass on the nematic phase.

Another point which deserves mention is the dependence of the surface energy
stored on the interface with the temperature. In this sense we have used a
mean field approximation for the determination of the correlation length,
namely $\xi \approx (T -T^*)^{-1/2}$, where $T^*$ is a temperature
for the structural phase transition in the smectic-nematic interface.
 From this observation and by
considering that $\varepsilon \approx \xi $ and $\beta \approx
K_2/\varepsilon $, 
 we can conclude that, near T* $\beta \approx K_2(T -T^*)^{1/2}$,
 where $K_2$ is assumed to be
temperature independent. This hypothesis is not valid near the nematic-isotropic
temperature transition ($T_{\rm NI}$), 
because in this case $K_2 \propto T_{\rm NI} -T$. Since we are close to the
smectic-nematic transition, the temperature dependence of $K_2$ can
be neglected.
This result indicates that as the transition temperature is approximated,
the surface energy of the interface becomes negligible. It also indicates 
that $\beta $ has a non-monotonic behavior with the temperature, which
agrees with the results obtained by Di Lisi et al. \cite{Lisi} for a
structural transition at a nematic-substrate interface.

Let us briefly discuss the main consequences of our model.
In dealing with the problem of the surface energy in
nematics, it is convenient to divide the contributions
to the anisotropic part of the surface energy into two parts: 
the extrinsic part, which comes  
directly from the nematic-substrate interaction, and the intrinsic one.
The intrinsic contribution comes, obviously, from the NLC 
itself. In a pseudo-molecular basis, it is usually
connected to the incomplete interaction among the liquid crystal
molecules near to the substrate, where the symmetry is reduced.
This introduces a spatial dependence on the elastic constants
in the vicinity of the surface. As suggested by 
Yokoyama et al.\cite{Yokoyama}
and reconsidered by 
many authors\cite{Faetti,Alexe,Durand,lre1}, 
the spatial variation of the elastic
constant can be considered as equivalent to a surface energy.
In a simple elastic model, based on the Maier-Saupe approximation,
the elastic contributions to the surface energy are connected to
the the spatial variation of the elastic constant and to the
spatial variation of the scalar order parameter\cite{Ponti}.
This later contribution is, in general, the dominant one,
because the extrapolation length connect to it is found to be
of the order of the coherence length in the NLC phase, and then
it is in the experimentally detectable range.  

Recently, 
an analysis of the anchoring competition between short-range and long-range
nematic-substrate interactions has been considered\cite{Barberi}.
In this approach, the alignment of a nematic sample  
results from the competition between an external, 
position dependent field, localized in a microscopic layer or a local
surface field, and the nematic-substrate interaction. 
It is possible to show that the phenomenon
presents a threshold behavior like the Fre\' edericksz transition. 
The analysis shows that according to the anchoring strength the final
orientation of the whole sample can be planar, distorted 
or, for large values of the anchoring strenght, homeotropic.
        
In our approach the competition effect is
not considered because the boundary conditions impose a perfect smectic
layer near the substrate and a nematic order in the bulk. 
Moreover, the long-range parts of the surface energy 
are not considered\cite{Petrov,Durand2,Local}.
The emphasis is
then on the effects along the interface where the orientation changes. 
Our model shows that, if this effect is present in a real nematic sample,
then the contribution of this interface to the surface energy has to be
taken into account. This is reinforced by the fact that the order of 
magnitude of this energy is comparable with the one experimentally
measured\cite{Blinov}. In fact, this order of magnitude of the contribution
to the surface energy depends on the thickness of the coherence length
in a crucial manner. As stressed before, 
for reasonable estimates of this quantity
it seems that this contribution is the dominant one. 
On the other hand it is well-known that the existence of 
geometrical non-uniformities can affect the surface energy 
experimentally
detected\cite{Berreman,Actual}.  
In fact, this kind of non-uniformity 
can be responsible for a surface energy which is of 
geometrical
origin and of elastic nature. In some cases of surface 
treatment or
surface shape, the 
experimental situation can lead one to detect an
apparent extrapolation length (i.e. an apparent anchoring energy)
when the expected situation is that of strong anchoring
(infinite anchoring strength)\cite{lre2}. 
Actually, a  
surface energy which is localized in the vicinity of the interface 
and
whose origin is connected with the geometrical properties of the
surfaces is a true
surface energy. It is difficult to distinguish between 
this contribution and the one which arises from the interaction 
of the NLC molecules and the substrate, i.e., the extrinsic
contribution.

Then, the above conclusions regarding our model for the
contribution of the smectic-nematic interface can be valid at least 
in the case in which perfectly flat
substrates are supposed to form the slab.

Consequently, among all the contributions to the surface energy,
the contribution arising from the smectic-nematic interface can not be
neglected. It has an
order of magnitude which  can be  
comparable with the one arising from the spatial variation of
the scalar order parameter and with those connected to the geometrical
effects. 

\acknowledgments

This work was completed during the visit of G. Barbero to our Department.
Many thanks are due to him for illuminating discussions. 

\end{document}